# The mesoscale order of nacreous pearls

Jiseok Gim[a], Alden Koch[b], Laura M. Otter[c], Benjamin H. Savitzky[d], Sveinung Erland[e], Lara A. Estroff[f,g], Dorrit E. Jacob[c], and Robert Hovden[a,h,1]

[a]Department of Materials Science & Engineering, University of Michigan, Ann Arbor, MI 48109; [b]Department of Biomedical Engineering, University of Michigan, Ann Arbor, MI 48109; [c]Research School of Earth Sciences, Australian National University, Canberra, ACT 2600, Australia; [d]National Center for Electron Microscopy, Molecular Foundry, Lawrence Berkeley National Laboratory, Berkeley, CA 94720; [e]Department of Maritime Studies, Western Norway University of Applied Sciences, 5528 Haugesund, Norway; [f]Department of Materials Science and Engineering, Cornell University, Ithaca, NY 14853; [g]Kavli Institute at Cornell for Nanoscale Science, Cornell University, Ithaca, NY 14853; and [h]Applied Physics Program, University of Michigan, Ann Arbor, MI 48109

A pearl's distinguished beauty and toughness are attributable to the periodic stacking of aragonite tablets known as nacre. Nacre has naturally occurring mesoscale periodicity that remarkably arises in the absence of discrete translational symmetry. Gleaning the inspiring biomineral design of a pearl requires quantifying its structural coherence and understanding the stochastic processes that influence formation. By characterizing the entire structure of pearls (~3 mm) in a cross-section at high resolution, we show that nacre has medium-range mesoscale periodicity. Self-correcting growth mechanisms actively remedy disorder and topological defects of the tablets and act as a countervailing process to long-range disorder. Nacre has a correlation length of roughly 16 tablets (~5.5 μm) despite persistent fluctuations and topological defects. For longer distances (>25 tablets, ~8.5 μm), the frequency spectrum of nacre tablets follows $f^{-1.5}$ behavior, suggesting that growth is coupled to external stochastic processes—a universality found across disparate natural phenomena, which now includes pearls.

nacre | pearls | TEM | SEM | mesoscale

The artist Jorge Méndez Blake illustrated a basic principle of structure in mesoscale periodicity and grown nanocrystalline materials when he placed a single book (i.e., a defect) in a brick wall, displacing every subsequent brick layer (1). Blake's work demonstrates the principle of paracrystallinity: disorder from any one defective site propagates throughout the layers of the material (2–4). In the context of brick laying, skilled masons overcome this difficulty using external templates to achieve translational order—marked guideposts and lacing cords align layers to prevent disorder from propagating (5, 6). Only with great calculated effort can a 10-story building ensure the same number of aligned brick layers on all sides (7). Regarding layered growth of nanomaterials, of course, external templates are limited and rare. Mesoscale periodicity—defined here as long-range translational order of mesoscopic building blocks—is thus improbable, due to natural variations in the unit sizes, without the aid of some additional countervailing mechanism. The rare instances where nature assembles mesoscale periodicity therefore merit our attention (8). Nacre in pearls and mollusk shells are one such example where mesoscale periodicity arises in an environment without discrete translational symmetry (9–11).

Pearls are renowned and coveted for their beauty; that beauty results from the diffracting iridescence of periodically stacked tablets (~500-nm-thick units) known as nacre (12–15). Nacre tablets form within interlamellar organic sheaths (~10- to 20-nm thickness) that act as the main patterning agent between layers. During formation, not-yet–calcified layers grow within multiple closely spaced sheaths that separate progressively during mineral growth (16, 17). While some work has focused on describing the atomic-to-nanoscale structure of the tablets themselves as assemblies of nanoparticles of aragonite (18–21) or the atomic order within and between tablets (22), the current work is focused on how these tablets are arranged at the next length scale (hundreds of nanometers—several microns [i.e., herein, the mesoscale]).

Mesoscale periodic nacre is a highly ductile structural composite that can withstand mechanical impact and exhibits high resilience on the macro- and nanoscale (23–25). Because of its superior toughness, nacreous shells protect the mollusk's soft body (11, 26–30) and inspire scientists designing next-generation supercomposites (31–33). Yet, despite a century of scientific fascination with nacre (9, 11, 24, 26, 34), its astonishing mesoscale periodicity has not been quantified, leaving key questions about this material unanswered: does nacre have long-range order? What is the stochastic process that governs its formation?

Here, we show nacre forms remarkable medium-range mesoscale periodicity through corrective processes that remedy disorder and topological defects. The entire nanostructure of nacreous pearls is characterized in cross-section to reveal complex stochastic processes that influence ordered nacre growth. Beginning atop an initially formed organic center, nanoparticles self-assemble into bulk aragonite, later followed by nacre deposition. The initial layers are disordered; however, this disorder attenuates within the first 200 layers through corrective growth processes that persist throughout the entire pearl. When a tablet is grown too thin, the next tends to be thicker—and vice

> **Significance**
>
> Despite a century of scientific fascination with nacre's periodic mesostructure, there remains a fundamental question of whether long-range order exists. In this work, the stochastic growth that leads to mesoscale order in the nacreous pearl is revealed by quantifying its structural coherence across entire pearl specimens. We find that mollusks strike a balance between preserving translational symmetry and minimizing thickness variation of layers by creating a paracrystal with medium-range order. Self-correcting growth processes allow pearls to quickly attenuate disorder, accommodate topological defects in tablet structure, and maintain order throughout a fluctuating external environment. These observations were made possible by characterizing the entire structure of Akoya "keshi" pearls (~3 mm) at high resolution (<~3 nm).

versa—thus compensating for the initial error. Local irregularities in thickness (±15° variation of interfacial curvature) and topological defects ($5.9 \times 10^8$ m$^{-2}$ defect density) intermittently appear, yet the pearl maintains medium-range order with a correlation length of ~16 tablets (~5.5 μm). At larger distances, the positions of tablet layers are uncorrelated. However, a pearl does not have perfect translational symmetry, and local disorder persists into subsequent layers—a paracrystalline property which ultimately limits long-range order. For longer-length scales (e.g., >25 tablets, ~8.5 μm) the aperiodic fluctuation of nacre's tablet thicknesses follow $f^{-1.5}$ noise behavior [$f = 1 / $ (number of tablets)], which could represent external Markov processes with longer memory attributable to cooperative environmental changes such as temperature, pH, food availability, seasonality, and tidal cycles.

## Results

**The Start of Nacre in Pearls.** Pearls form either as a natural response to mantle tissue injury or when mantle tissue is deliberately transplanted from a donor into a host animal for pearl culturing. In both cases, the mantle epithelium develops a closed cyst—the so-called pearl sac—programmed to reproduce the structure of the shell (35). CaCO$_3$ is then secreted onto a manufactured bead (for bead-cultured pearls) or any available debris enclosed by the developing pearl sac (for nonbeaded pearls) within a limited space (36). The shape of the manufactured or natural nucleus usually dictates a pearl's macroscopic shape (36, 37) (*SI Appendix*, Figs. S1–S5). Here, we primarily study non–bead-cultured "keshi" pearls, grown for ca. 18 mo in *Pinctada imbricata fucata* mollusks on the Eastern shoreline of Australia (*Methods*) (38).

The cross-section of a nonbead "keshi" pearl is shown in Fig. 1B. The mollusk deposits calcium carbonate upon the irregular organic center (~200-μm diameter) and after roughly ~15 μm of bulk aragonite, nacre layers begin to form. Nacre layers are bonded by chitin-proteinaceous organic sheaths periodically secreted progressively in the direction toward the margin by the epithelium (39–41). Ordered nacre makes up 87% of this 2.5-mm pearl (Fig. 1 B and C). However, the nacre is preceded by more disordered growth stages. Nacre and other biominerals form by nonclassical crystallization via metastable transient precursor phases (42, 43) and attachment of CaCO$_3$ nanoparticles interspersed by organic macromolecules (Fig. 1 *G* and *H* and *SI Appendix*, Fig. S6). The nanoparticle packing intermittently increases until forming columnar aragonite with segregated regions of organic. In Fig. 1*B*, this occurs over roughly 200 μm of growth. Throughout this process, the pearl's volume becomes larger and rounder, before penultimately forming a bulk aragonite (Fig. 1*C*). In most observations, we observe bulk aragonite begins after a region reaches positive curvature (44–46).

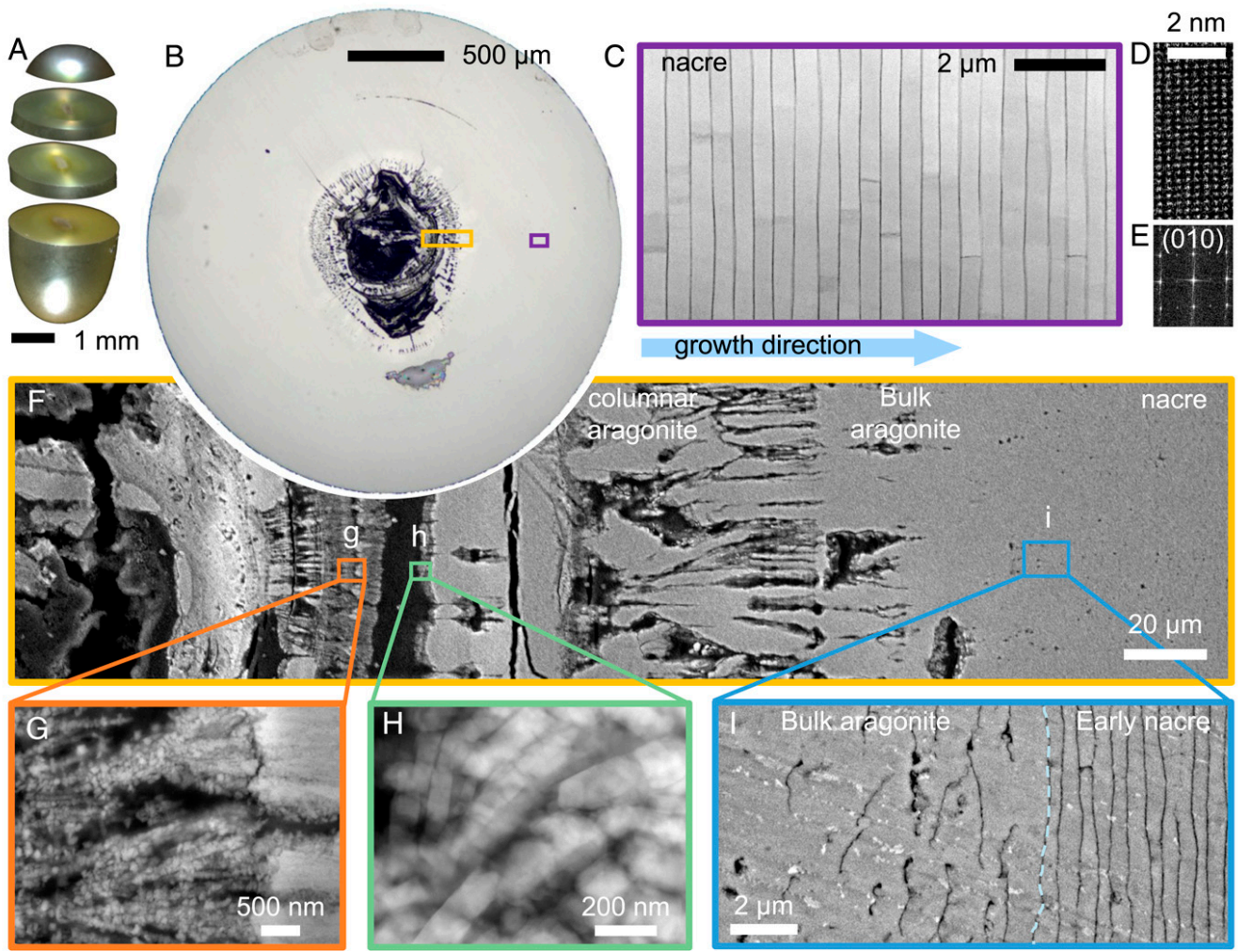

**Fig. 1.** Formation of non–bead-cultured Akoya "keshi" pearls produced in a *P. i. fucata* mollusk. (*A*) Optical overview of a nonbeaded keshi–cultured pearl showing iridescence due to the interplay of constructive interference at nacre tablet layers with light. (*B*) Cross-section showing CaCO$_3$ growth begins onto an organic center. (*C*) Mature nacre (purple box in *B*) showing the ordered state in their thickness and interface curvature. (*D* and *E*) Atomic-resolution ADF-STEM of mature nacre and its corresponding Fourier transform indicating highly crystalline nacre and a lattice constant that is consistent with aragonite. (*F*) Cross-sectional backscatter SEM at the center of the pearl (yellow box in *B*) showing transition from spherulitic aragonite structures to nacre. (*G* and *H*) Aggregated nanoparticles (at time of observation) form massive, structurally indistinct aragonite structure. (*I*) Formation of nacre begins directly on massive, structurally indistinct aragonite.

In pearls, nacre forms abruptly on bulk aragonite. At the start of nacre, around 440 μm (±12%) from the organic center in Fig. 1, the bulk aragonite-to-nacre interface is approximately flat (curvature of ∼0.002 μm$^{-1}$) with locally rough texture (∼200-nm variation over 7 μm). About 10 to 20 μm prior to the first nacre layer, larger organic deposits appear in comparable quantity to the interlayer sheaths that appear in mature nacre—suggesting the genetic processes directing the system toward sheath deposition start earlier than the structure itself (*SI Appendix*, Fig. S9). This same transition observed in *Pinctada fucata* pearls was also found in nonbead cultured Tahitian "keshi" pearls produced by the *Pinctada margaritifera* mollusk (*SI Appendix*, Fig. S8). This direct nacre growth onto bulk aragonite in pearls is reminiscent of *P. fucata, Unio pictorum,* and *Nautilus pompilius* shells but is distinct from nacre formation reported in *Pinna nobilis*, where the assembly process is driven by aggregation of nanoparticles within a several-micrometers-thick organic matrix (11, 47, 48).

**Mesoscale Periodicity of Nacre.** The first nacre tablets are non-uniform, with substantial thickness variation and a rough interface with the preceding bulk aragonite. The Fourier transform of early nacre layers imaged by backscattered electron scanning electron microscopy (BSE-SEM) reveals ±15° angular variation in their interface curvature (Fig. 2*A*, *Inset*). This angular broadening decreases quickly to ±6° after 100 layers and further down to ±5° after 200 layers (mature nacre). This smoothing of tablet interfaces is clearly visible (Fig. 2*A* and *SI Appendix*, Fig. S10). At the same time, tablet thickness decreases by 30%, reducing from 500 ± 300 nm for early nacre to 340 ± 120 nm (a ∼40% reduction of variance) for mature nacre. In mature nacre (Fig. 2*A*), peaks visible in the Fourier transform (first peak at (∼340 nm)$^{-1}$) provide the signature presence of mesoscale periodicity (i.e., translational symmetry of tablets). However, these peaks in Fourier space do not quantify the long-range order nor describe the stochastic processes that create the mesoscale periodicity.

Pair-correlation functions quantify the mesoscale periodicity of nacre in pearls along the growth direction (Fig. 2*B*) by measuring the probability of tablet spacings (*Methods*). Subsequent peaks indicate the distance to the first nearest neighbor (NN), second NN, third NN, etc. These plots were calculated based on all pairs of tablets in each stage of nacre (Fig. 2*A*). The first peak represents the average distance between adjacent tablets, and subsequent peaks describe longer-range periodic order.

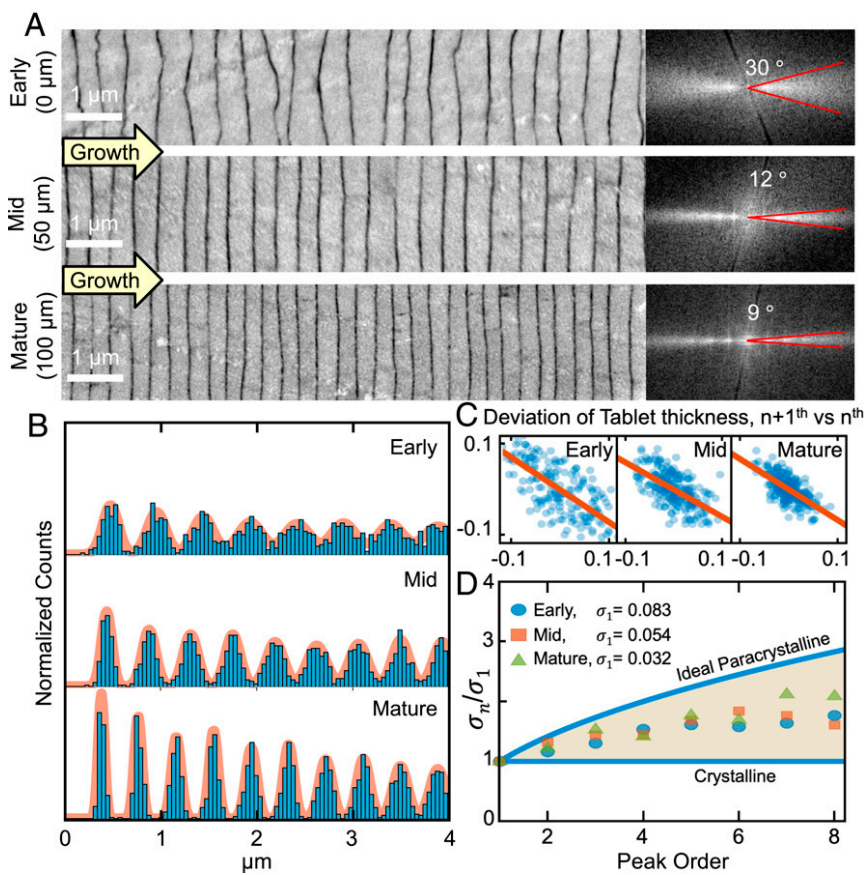

**Fig. 2.** Quantification of mesoscale periodicity in nacre. (*A*) Cross-sectional backscatter SEM of the early, middle, and mature stages of nacre growth shows ordering through reduced variation in the interface curvature and tablet thickness. Fourier transforms (*Right*) of nacre (*Left*) imaged by BSE-SEM show that angular broadening decreases from ±15 to ±5°. (*B*) Pair-correlation functions of nacreous layers represent the probability of finding tablets spaced a given number of unit cells apart. Sharpening of peaks in later nacre indicates increasing long-range order. (*C*) Correlation of the thickness of tablets with nearest neighbors shows a negative correlation. If one tablet is thick, the next one tends to be thin. (*D*) Cumulative disorder in nacre described by a real paracrystalline model, demonstrating that nacre has order between that of a crystal and a paracrystal model.

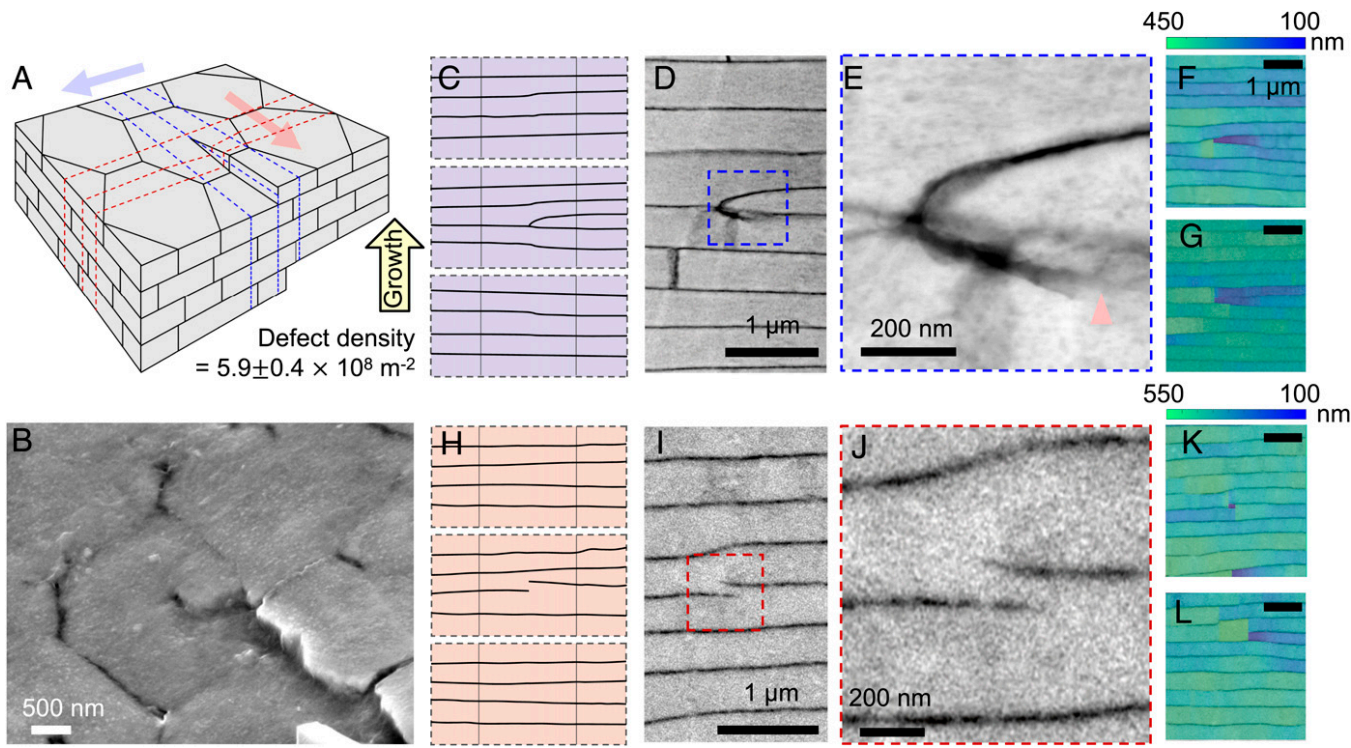

**Fig. 3.** Topological defects in nacre. (*A*) Schematic of topological defects (a screw dislocation) in nacre. (*B*) Forty-five-degree tilted backscatter electron microscopy showing defect. (*C*) Cross-section of the schematic perpendicular to the slip plane (along direction of blue arrow in *A*). (*D* and *E*) ADF-STEM showing the extra tablet generated due to a topological defect and the mineral bridge connected to the adjacent layer. (*F* and *G*) Thickness map of the extra tablet showing abrupt change of thickness at the point of defect. (*H*) Cross-section of the schematic along the slip plane (along the direction of red arrow in *A*). (*I* and *J*) ADF-STEM showing the extra organic interface split by the defect. (*K* and *L*) Thickness map of the extra organic interface showing an abrupt change of thickness at the point of defect.

For ideal crystals, every peak will appear equally sharp. However in nacre, peaks in the pair-correlation function increasingly broaden—indicating propagation of disorder, as described by the paracrystal model (4, 49, 50).

Nacre is well-described by an ordered paracrystal or a medium-ranged mesoscale crystal model. We report mature nacre has a correlation length of 5.5 μm, and translational order is lost by 16 layers. Correlation length is defined as the length where the pair-correlation envelope disappears (see *Methods* and *SI Appendix*, Fig. S11). Thus, although nacre maintains extraordinary translational symmetry across a dozen or so layers, it lacks true long-range order. In paracrystals, a deviation in layer thickness displaces subsequent layers and thus disorder propagates. Defects and tablet variation degrade translational symmetry with distance. The progressive broadening of each peak in the pair-correlation functions quantifies the long-range order and paracrystallinity (Fig. 2*D*). In a perfect paracrystal, the SD of peaks, $\sigma_n$, broadens by the square root of its peak number ($\sigma_n = \sigma_1\sqrt{n}$) (4, 51).

Nacre, at the mesoscale, has both crystalline-like and paracrystalline disorder. As shown in Fig. 2*D*, nacre falls below the paracrystal curve (more ordered) and above the flat crystalline curve. The paracrystallinity, or broadening of peaks, is approximately equivalent in early, mid, and mature-stage nacre, suggesting the growth mechanisms may be equivalent. However, the sharper first-order peaks in mature nacre reflect improved initial conditions.

Nacre maintains discrete translational symmetry through self-correcting growth processes. The bumps and valleys in one nacre layer are attenuated in the subsequent layers. If one tablet is thicker than average, the next tends to be thinner. This negative correlation of thickness between adjacent nacre layers is quantified in the early, mid, and mature nacre in Fig. 2*C*. The negative correlation ($-0.66$, $R^2 = 0.46$, $P = 10^{-24}$) follows an autoregressive model that allows ordered periodic growth to reach steady state in ∼200 layers, which demarcates maturity. In uncorrelated growth, thickness variation of each nacre tablet would devastate the periodic long-range order—much like Jorge Méndez Blake's brick wall (1). Unlike a brick wall, multiple nacre layers grow simultaneously within continuous organic sheaths. Ultimately, nacre is able to self-correct for fluctuations in tablet thicknesses and occasional screw dislocations (Fig. 3 *A* and *B* and *SI Appendix*, Figs. S12 and S13).

**Topological Defects in Nacre.** Local disorder in pearls encompasses topological defects (i.e., dislocations) of the tablets. SEM images at the pearl surface and a corresponding schematic illustrate how screw dislocations originate in nacre (Fig. 3 *A* and *B* and *SI Appendix*, Figs. S12 and S13). High-resolution annular dark-field scanning transmission electron microscopy (ADF-STEM) images show a dislocation in cross-section (Fig. 3 *C*–*L* and *SI Appendix*, Fig. S13). Screw dislocations in nacre are known (52–54) and recently reported to couple as chiral pairs through a dissipative distortion field that helps accommodate the space filling requirements of nacre (34). In cross-sections, screw dislocations can be misidentified as edge dislocations (although both may be present). When viewed parallel to the slip plane, a mineral bridge at the dislocation marks the defect origin. Viewed normal to the slip plane the screw dislocation appears as an additional partial organic boundary. The defect density (54) of nacreous pearls herein is $5.3 \pm 0.4 \times 10^8$ m$^{-2}$ in plane view and $6.5 \pm 0.4 \times 10^8$ m$^{-2}$ in cross-section (Fig. 3*A* and *SI Appendix*, Fig. S14). Our quantification is measured from top view and cross-sectional SEM by dividing the total number of defects by the total area (*SI Appendix*, Fig. S14).

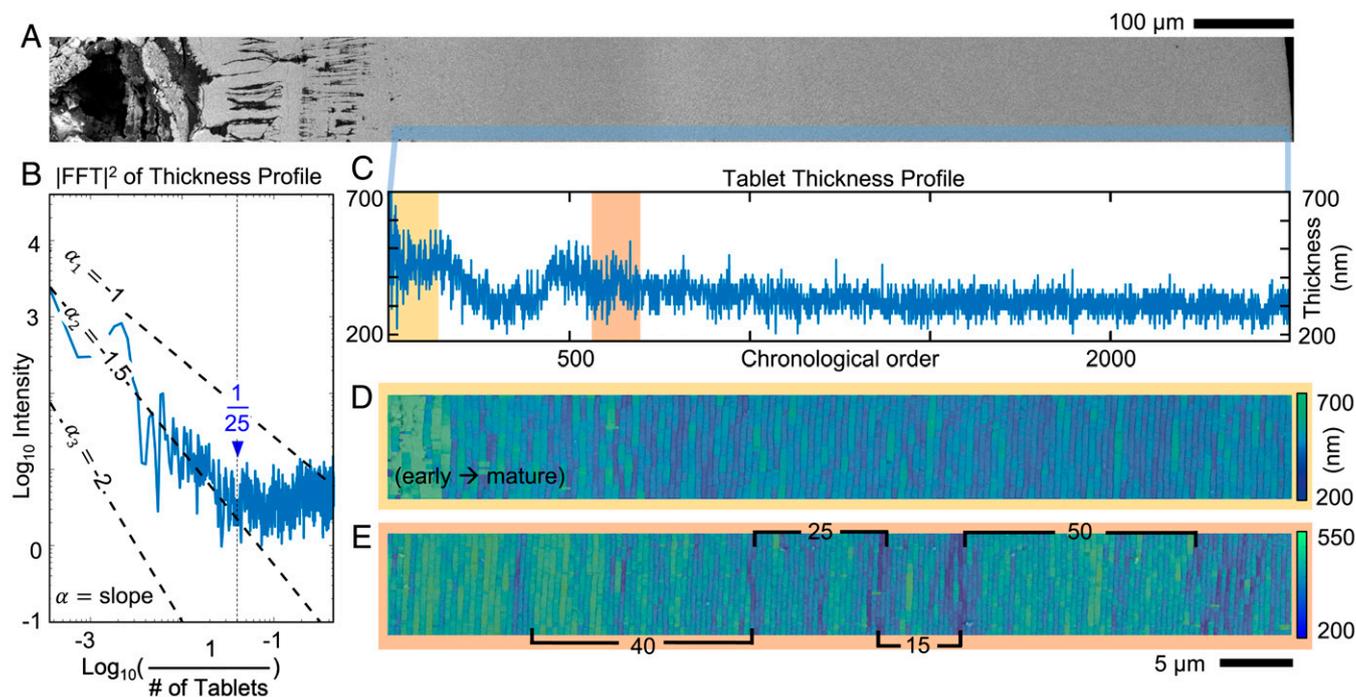

**Fig. 4.** Growth processes of nacre throughout the entire cross-section of the pearl. (*A*) Overview of the pearl cross-section spanning from center to edge. (*B*) Log–log plot of the spectral density of the tablet thickness profile showing nacre thickness variation described by Markov processes. (*C*) Thickness profile across the entire pearl cross-section. (*D*) Thickness map from the early to mature stage of nacre corresponding to early nacre. It shows an abrupt attenuation of disorder in thickness and interface curvature. (*E*) Thickness map of the mature nacre showing different length scales associated with tablet thickness fluctuations.

Despite the prevalence of dislocations, nacre's self-correcting growth processes preserve mesoscale translational symmetry via substantial thickness changes (∼28%) at the topological singularity. Fig. 3 *F*, *G*, *K*, and *L* show tablet thicknesses at and around screw dislocations viewed in cross-section parallel or perpendicular to the slip plane. The singularity causes tablets to become thinner to accommodate an additional layer without disrupting the long-range order of the mesoscale periodicity. Within just a couple layers, above or below the dislocation, nacre visually returns to uniform, periodic growth.

**Nacre Growth Fluctuations Follow *1/f* Behavior.** To understand the stochastic growth processes over longer lengths and time, nacre tablet thicknesses are analyzed across the entire pearl (2.5-mm diameter) at high resolution (∼3 nm) (Fig. 4*A*). This pearl contained 2,615 tablet layers deposited over 548 d, giving a mean growth rate of 4 to 5 tablets (1.4 to 1.7 μm) per day—in this regard, distances may also be considered in terms of time (*Methods*). Every tablet of nacre is measured from a mosaic of BSE-SEM images, and a profile of their thicknesses is plotted in chronological order in Fig. 4*C*. Initial thicknesses fluctuate substantially but reduce quickly within ∼10 layers (Fig. 4*C*, yellow box), as shown in the thickness map (Fig. 4*D*). Notably, sawtooth-shaped bursts in systematic thickness variation are visible (Fig. 4*C*, red box) and repeat aperiodically throughout the entire growth (Fig. 4*E*). The stochastic processes of nacre growth are revealed in a power spectral density (PSD [i.e., squared magnitude of the Fourier transform]) of the thickness profile (Fig. 4*B*). The spectral density shows two distinct regimes: power-law decay across a low- to mid-frequency regime corresponding to distances longer than ∼25 tablets and a high-frequency regime where the density slightly increases.

The spectral density of a pearl's tablet thicknesses follows power-law decay across low to mid frequencies, colloquially called *1/f* noise, suggesting nacre growth has correlations that extend over a wide range of time scales with cooperative effects linked to environmental changes. In stochastic Markov processes, the distribution of future events are determined by recent past events (55, 56), and even simple Markov processes can exhibit a spectral density following $f^{-\alpha}$ where $1 < \alpha < 2$. This behavior characterizes a wide range of phenomena from fluctuations in river flow to heartbeats or economic markets (57, 58). The spectral density of nacre follows $f^{-1.5}$ across low to mid frequencies (>∼25 tablets)—here, $f = 1/$ (number of tablets). In real-space, this corresponds to the sawtooth-shaped bursts in the thickness profile of Fig. 4*C*. Such aperiodic fluctuations can be associated with the $f^{-1.5}$ spectral density, and the behavior can be explained by the additive effect of one or several Markov processes representing the external effect of environmental and physiological factors. The coefficient α of power-law decay is estimated by truncating the PSD and applying a square fit for log–log coordinates (*Methods*).

The observed power-law decay, $f^{-1.5}$, suggests that nacre's growth is influenced by a set of stochastic processes with different characteristic time scales. This makes sense: we expect growth to depend on many external factors in the surrounding habitat of the mollusk shell (13, 14, 59). Multiple clocks in nature govern these living animals, such as the 30-d circalunar cycle (60, 61), the day–night circadian cycle (62, 63), tidal changes (64), hibernation in the winter (65, 66), and aestivation in the summer (67, 68). This fluctuating marine environment and the natural time scales of the mollusk are interconnected influences of the nonequilibrium thermodynamical conditions (69–71) of nacre growth and likely explains the observed power-law behavior.

At high frequencies (<∼25 tablets), there is a flattening of the *1/f* noise and a slight increase in the power spectrum due to

Gim et al.
The mesoscale order of nacreous pearls

PNAS | 5 of 8
https://doi.org/10.1073/pnas.2107477118Downloaded by guest on October 21, 2021

corrective growth processes, whereby the thickness of each tablet is anticorrelated with the thickness of the previous layer. The opposing fluctuations in subsequent layers (shown in Fig. 2C) amplifies high-frequency variations in a manner consistent with a negative autoregressive process (72–74) and causes the slightly increasing profile of the spectral density at high frequencies (Fig. 4B). Mechanistically, the source of this correlative correction is not clear—it may be explained by a partial viscosity of the $CaCO_3$ or the precursor phase upon deposition under hydrostatic pressure. The corrective behavior likely emanates when nacre tablets form simultaneously within multiple periodically spaced continuous organic sheaths—allowing not-yet–calcified layers to influence the thicknesses of adjacent layers. Physiological responses may also play a role.

In nacre, we see a fundamental tradeoff between achieving mesoscale translational symmetry and minimizing tablet thickness variation. Corrective growth occurs to counterbalance disorder and the natural tendency toward paracrystallinity, thereby enabling surprising and remarkable mesoscale periodic ordering. However, this corrective growth process amplifies variation, wherein perturbations in one layer causes an opposite perturbation in subsequent layers.

## Discussion

Pearls are iconic mesoscale periodic structures grown by mollusks with no external templates along the direction of growth. Nanoparticles assemble within an organic matrix with increasing density until a massive, bulk aragonite forms and later followed by nacre deposition. Tablets are then grown with self-correcting mechanisms that allow mesoscale translational symmetry to accommodate tablet variation and topological defects. Mollusks strike a balance between preserving translational symmetry and reducing thickness variation by creating a paracrystal with medium-range order (∼5.5-μm correlation length). This balance allows pearls to attenuate the initial disorder during early formation and maintain order throughout a changing external environment. Over longer-length scales (>∼25 tablets), nacre growth is mediated by external stochastic processes and exhibits a "$1/f$" behavior that belongs to a universal class of vast and disparate phenomena. Quantifying the mesoscale periodicity and paracrystallinity of nacreous pearls reveals underlying processes of formation and illuminates metrics for assessing assemblies of synthetic nanocrystalline materials that seek to mimic nacre's hierarchical design.

## Methods

**Specimen.** Specimens of nacreous non–bead-cultured Akoya "keshi" pearls produced by the *P. i. fucata* pearl-oyster were collected at the Broken Bay Pearl farm (Pearls of Australia Pty Ltd) located at the East coast of Australia. The term keshi pearls is used today to describe beadless pearls that were unintentionally produced as a byproduct of culturing bead-cultured pearls (75). This farm was chosen as it avoids postharvest treatments such as bleaching or dying as is otherwise common practice (38).

**Wedge-Polishing Preparation.** Cross-sections were cut from the whole pearls measuring 3 to 5 mm in diameter using a diamond wire saw. The cross-sections for S/TEM were prepared by mechanical wedge polishing (11, 24), which provides large-area, electron-transparent specimens with structural stability. The lapping process was performed using an Allied MultiPrep polishing system and a series of micrometer-sized diamonds embedded in polishing film (from 30 μm down to 0.1 μm). The lapping sequence was dictated by the "trinity of damage," where a damaging layer is assumed to be less than three times as thick as the size of the grit. Therefore, with each subsequent lapping film, this damage layer was removed. To prevent $CaCO_3$ etching during preparation, water-based lubricants were eliminated—instead, Allied alcohol-based "blue" lubricant was used.

**Electron Microscopy.** SEM images were recorded with a Z-contrast backscattered electron detector of a JEOL IT500 system (10 kV) with an energy-dispersive X-ray spectroscopy detector. High-angle ADF-STEM was performed using a JEOL 3100R05 microscope with Cs aberration-corrected STEM (300 keV, 15 mrad) and cold field emission gun. An ADF detector with 120- to 150-mm camera lengths and a detector angle of 59 ∼ 74 (inner) to 354 ∼ 443 mrad (outer) were used to produce Z-contrast images, where greyscale intensity is sensitive to the atomic number in the specimen's matrix. Column pressure in the STEM column at the specimen was ∼$1 \times 10^7$ torr.

Low-dose methods, beam shuttering, and examination of regions exposed to the beam were used to separate electron beam irradiation from intrinsic phenomena (11, 24). For STEM measurements, with typical fields of view from ∼500 nm to 10 μm, the electron dose was typically from ∼0.4 to 90 $e^-·Å^{-2}$, and dose rates ranging from around ∼0.1 to 2.7 $e^-·Å^{-2}·s^{-1}$. The material was structurally preserved during imaging. However for the same imaging conditions at higher magnifications (e.g., 20-nm field of view), the radiation dose increases to ∼$9 \times 10^5$ $e^-·Å^{-2}$ and dose rate to ∼$3 \times 10^3$ $e^-·Å^{-2}·s^{-1}$, which causes the material to show electron irradiation damage localized to the small field of view. Thus, larger fields of view are preferred to minimize dose and provide a large area of observation. Atomic-resolution STEM requires a small field of view with an on-axis region of interest.

Nacre's growth rate is counted based on the age of pearl. We assume that only nacre growth takes out most of the pearl's lifetime because nacre occupies almost 90% of the entire volume compared to the bulk aragonite, and the growth rate of nacre along c-axis is slower than that of the aragonite (76, 77). Thus, the growth rate of nacre is the total number of nacre layers (e.g., 2,615 layers) divided by the lifetime of pearl (e.g., 18 mo ∼ 548 d).

**Raman Spectroscopy.** Confocal Raman spectroscopy (*SI Appendix*, Figs. S7 and S8) was performed on the cross-sections of pearls using a Tescan Raman imaging and scanning electron microscope equipped with a WITec confocal Raman microscope. Raman spectra in the spectral range (100 to 3,700 $cm^{-1}$) were recorded with a charge-coupled device camera using a solid-state 532-nm laser as the excitation source (spectral grating: 600 g/mm, laser power: 10 ∼ 15 mW, optical lens: 100×). Each Raman spectrum was acquired typically for 0.5 ∼ 1 s, and with 30 ∼ 100 of scans to minimize noise effects. The parameters for excitation power, acquisitions and accumulations are selected to prevent the sample from laser burn. Spectral reproducibility was confirmed by taking several spot analyses.

**Data Analysis.** The pair-correlation function describes the probability distributions of NN tablets in space about a known tablet position (50). The one-dimensional pair-correlation function is extracted from raw BSE-SEM images with high resolution (∼3 nm). Mechanical scratches are removed using the image filtering (78). Interlamellar sheaths are identified by finding local minima in BSE intensity line profiles taken perpendicular to the nacre layers. The resulting interfacial positions are then used to calculate the thickness of nacre layers by measuring the distance between positions. The $N^{th}$ tablet thickness is calculated as the difference between the $N$th and $(N+1)$th interface positions. This is done throughout all positions and layers across the entire pearl cross-section to generate a thickness map visualized as a color overlay atop an SEM image. A one-dimensional (1D) histogram of tablet thickness was then calculated and normalized by a factor of $\frac{1}{k*bw}$ to obtain a discretization of the pair-correlation function, where $k$ is the number of samples and $bw$ is the histogram bin width. A sum of normalized gaussians is fit to the histogram peaks, and the SDs are plotted versus peak order. Occasional, sudden, narrow two-tablet thicknesses appear in the thickness maps due to lack of contrast between layers, which may be due to topological defects and sometimes large asperities or mineral "Checa bridges" between tablets (20)—these minority regions were not included in generating the pair-correlation function.

Correlation between the widths of adjacent tablets is calculated by measuring the deviation in tablet width from the local mean tablet width. The local mean is calculated from the list of all sampled tablet widths using a rolling average with a width of three tablets. Thus, the deviation in thickness of the $n$th tablet is given as $width_N - (width_{(N-1)} + width_N + width_{(N+1)})/3$. Deviations in thickness of pairs of adjacent tablets are plotted against each other and the correlation coefficient is estimated by linear least squares fitting. Correlation length is defined as the length where the pair-correlation envelope disappears. It becomes negligible when the magnitude of the following peak is below 10% of the first peak (i.e., $\frac{h_x}{h_1} < 0.1$). Transitional symmetry is lost by 16 layers (i.e., 5.5 μm).

Autocorrelation of tablet thicknesses can be interpreted by PSD, which is generated in the frequency domain by squaring a fast Fourier transform of thickness profile. The tablet thickness is measured from BSE-SEM images using the peak-detection algorithm described above and is plotted in chronological order, where the values are evenly spaced. Since the Fourier transform is not a consistent estimator (the variance is not reduced when the number of data points increases), the PSD is averaged over the profiles from 10 neighboring regions and is linearized in log–log scale to reveal 1/f noise. The logarithmic PSD is fitted by polynomial least squares methods to determine the power-law decay coefficient $\alpha$ of the 1/f noise component (i.e., the slope of 1/f noise) for $0.04 < f < 0.0004$ (25 to 2,500 tablets). For frequencies higher than 0.04, the 1/f noise is flattened and change the slope to be positive.

**Data Availability.** All study data are included in the article and/or SI Appendix.

**ACKNOWLEDGMENTS.** We acknowledge the University of Michigan College of Engineering for financial support and the Michigan Center for Materials Characterization for use of the instruments. We thank Pearls of Australia Pty Ltd and the Broken Bay Pearl Farm for providing the Akoya keshi pearls and Professor Pupa U. P. A. Gilbert for providing non–bead-cultured Tahiti keshi pearl. D.E.J. and L.M.O. are supported by Australian Research Council Grant Nos. DP160102081 and DP210101268. B.H.S. was supported by the Toyota Research Institute. We thank the reviewers for their careful feedback to improve the manuscript.

1. J. M. Blake, *The Castle* (José Cornejo Franco Library, Guadalajara, Mexico, 2007).
2. T. Bellini, L. Radzihovsky, J. Toner, N. A. Clark, Universality and scaling in the disordering of a smectic liquid crystal. *Science* **294**, 1074–1079 (2001).
3. D. A. Keen, A. L. Goodwin, The crystallography of correlated disorder. *Nature* **521**, 303–309 (2015).
4. R. Hosemann, W. Vogel, D. Weick, F. J. Baltá-Calleja, Novel aspects of the real paracrystal. *Acta Crystallogr. A* **37**, 85–91 (1981).
5. J. S. Wilson, Stability of structures. *Nature* **136**, 568–571 (1935).
6. A. W. Hendry, "Structural design of masonry buildings" in *Structural Masonry*, A. W. Hendry, Ed. (Macmillan Education UK, London, 1998), pp. 1–15.
7. R. Mark, P. Hutchinson, On the structure of the Roman pantheon. *Art Bull.* **68**, 24–34 (1986).
8. G. M. Whitesides, B. Grzybowski, Self-assembly at all scales. *Science* **295**, 2418–2421 (2002).
9. B. Bayerlein et al., Self-similar mesostructure evolution of the growing mollusc shell reminiscent of thermodynamically driven grain growth. *Nat. Mater.* **13**, 1102–1107 (2014).
10. G.-T. Zhou, Q.-Z. Yao, J. Ni, G. Jin, Formation of aragonite mesocrystals and implication for biomineralization. *Am. Mineral.* **94**, 293–302 (2009).
11. R. Hovden et al., Nanoscale assembly processes revealed in the nacroprismatic transition zone of *Pinna nobilis* mollusc shells. *Nat. Commun.* **6**, 10097 (2015).
12. L. Addadi, S. Weiner, A pavement of pearl. *Nature* **389**, 912–913 (1997).
13. P. U. P. A. Gilbert et al., Nacre tablet thickness records formation temperature in modern and fossil shells. *Earth Planet. Sci. Lett.* **460**, 281–292 (2017).
14. I. C. Olson, R. Kozdon, J. W. Valley, P. U. P. A. Gilbert, Mollusk shell nacre ultrastructure correlates with environmental temperature and pressure. *J. Am. Chem. Soc.* **134**, 7351–7358 (2012).
15. J. Salman et al., Hyperspectral interference tomography of nacre. *Proc. Natl. Acad. Sci. U.S.A.* **118**, e2023623118 (2021).
16. H. Nakahara, "Mechanisms and phylogeny of mineralization in biological systems," in *Nacre Formation in Bivalve and Gastropod Molluscs*, S. Sugea, H. Nakahara, Eds. (Springer, Tokyo, Japan, 1991), pp. 343–350.
17. J. H. E. Cartwright, A. G. Checa, The dynamics of nacre self-assembly. *J. R. Soc. Interface* **4**, 491–504 (2007).
18. J. Seto et al., Structure-property relationships of a biological mesocrystal in the adult sea urchin spine. *Proc. Natl. Acad. Sci. U.S.A.* **109**, 3699–3704 (2012).
19. Y.-Y. Kim et al., A critical analysis of calcium carbonate mesocrystals. *Nat. Commun.* **5**, 4341 (2014).
20. I. C. Olson et al., Crystal nucleation and near-epitaxial growth in nacre. *J. Struct. Biol.* **184**, 454–463 (2013).
21. H.-C. Loh et al., Nacre toughening due to cooperative plastic deformation of stacks of co-oriented aragonite platelets. *Commun. Mater.* **1**, 1–10 (2020).
22. P. U. P. A. Gilbert, M. Abrecht, B. H. Frazer, The organic-mineral interface in biominerals. *Rev. Mineral. Geochem.* **59**, 157–185 (2005).
23. F. Barthelat, H. Tang, P. Zavattieri, C. Li, H. Espinosa, On the mechanics of mother-of-pearl: A key feature in the material hierarchical structure. *J. Mech. Phys. Solids* **55**, 306–337 (2007).
24. J. Gim et al., Nanoscale deformation mechanics reveal resilience in nacre of *Pinna nobilis* shell. *Nat. Commun.* **10**, 4822 (2019).
25. H. D. Espinosa, J. E. Rim, F. Barthelat, M. J. Buehler, Merger of structure and material in nacre and bone – Perspectives on de novo biomimetic materials. *Prog. Mater. Sci.* **54**, 1059–1100 (2009).
26. F. Barthelat, Z. Yin, M. J. Buehler, Structure and mechanics of interfaces in biological materials. *Nat. Rev. Mater.* **1**, 1–16 (2016).
27. U. G. K. Wegst, H. Bai, E. Saiz, A. P. Tomsia, R. O. Ritchie, Bioinspired structural materials. *Nat. Mater.* **14**, 23–36 (2015).
28. Z. Deng et al., Strategies for simultaneous strengthening and toughening via nanoscopic intracrystalline defects in a biogenic ceramic. *Nat. Commun.* **11**, 5678 (2020).
29. H. D. Espinosa et al., Tablet-level origin of toughening in abalone shells and translation to synthetic composite materials. *Nat. Commun.* **2**, 173 (2011).
30. D. Grégoire, O. Loh, A. Juster, H. D. Espinosa, In-situ AFM experiments with discontinuous DIC applied to damage identification in biomaterials. *Exp. Mech.* **51**, 591–607 (2011).
31. E. Munch et al., Tough, bio-inspired hybrid materials. *Science* **322**, 1516–1520 (2008).
32. F. Bouville et al., Strong, tough and stiff bioinspired ceramics from brittle constituents. *Nat. Mater.* **13**, 508–514 (2014).
33. C. Zhao et al., Layered nanocomposites by shear-flow-induced alignment of nanosheets. *Nature* **580**, 210–215 (2020).
34. M. Beliaev, D. Zöllner, A. Pacureanu, P. Zaslansky, I. Zlotnikov, Dynamics of topological defects and structural synchronization in a forming periodic tissue. *Nat. Phys.* **17**, 410–415 (2021).
35. D. E. Jacob et al., Nanostructure, composition and mechanisms of bivalve shell growth. *Geochim. Cosmochim. Acta* **72**, 5401–5415 (2008).
36. L. M. Otter, U. Wehrmeister, F. Enzmann, M. Wolf, D. E. Jacob, A look inside a remarkably large beaded South Sea cultured pearl. *Gems Gemol.* **50**, 58–62 (2014).
37. M. Krzemnicki, D. Friess, P. Chalus, H. Hänni, S. Karampelas, X-ray computed microtomography: Distinguishing natural pearls from beaded and non-beaded cultured pearls. *Gems Gemol.* **46**, 128 (2010).
38. L. M. Otter, B. A. A. Oluwatooshin, L. T.-T. Huong, T. Hager, D. E. Jacob, Akoya cultured pearl farming in Eastern Australia. *Gems Gemol.* **53**, 423–437 (2017).
39. A. G. Checa, Physical and biological determinants of the fabrication of molluscan shell microstructures. *Front. Mar. Sci.* **5**, 353 (2018).
40. M. A. Meyers et al., The role of organic intertile layer in abalone nacre. *Mater. Sci. Eng. C* **29**, 2398–2410 (2009).
41. Y. Levi-Kalisman, G. Falini, L. Addadi, S. Weiner, Structure of the nacreous organic matrix of a bivalve mollusk shell examined in the hydrated state using cryo-TEM. *J. Struct. Biol.* **135**, 8–17 (2001).
42. J. J. De Yoreo et al., Crystallization by particle attachment in synthetic, biogenic, and geologic environments. *Science* **349**, aaa6760 (2015).
43. A. Gal et al., Particle accretion mechanism underlies biological crystal growth from an amorphous precursor phase. *Adv. Funct. Mater.* **24**, 5420–5426 (2014).
44. V. Schoeppler et al., Crystal growth kinetics as an architectural constraint on the evolution of molluscan shells. *Proc. Natl. Acad. Sci. U.S.A.* **116**, 20388–20397 (2019).
45. M. G. Willinger, A. G. Checa, J. T. Bonarski, M. Faryna, K. Berent, Biogenic crystallographically continuous aragonite helices: The microstructure of the planktonic gastropod *Cuvierina*. *Adv. Funct. Mater.* **26**, 553–561 (2016).
46. A. Freer, D. Greenwood, P. Chung, C. L. Pannell, M. Cusack, Aragonite prism−nacre interface in freshwater mussels *Anodonta anatina* (Linnaeus, 1758) and *Anodonta cygnea* (L. 1758). *Cryst. Growth Des.* **10**, 344–347 (2010).
47. L. Gránásy et al., Phase-field modeling of biomineralization in mollusks and corals: Microstructure vs formation mechanism. *JACS Au* **1**, 1014–1033 (2021).
48. K. Saruwatari, T. Matsui, H. Mukai, H. Nagasawa, T. Kogure, Nucleation and growth of aragonite crystals at the growth front of nacres in pearl oyster, *Pinctada fucata*. *Biomaterials* **30**, 3028–3034 (2009).
49. P. M. Voyles et al., Structure and physical properties of paracrystalline atomistic models of amorphous silicon. *J. Appl. Phys.* **90**, 4437–4451 (2001).
50. B. H. Savitzky et al., Propagation of structural disorder in epitaxially connected quantum dot solids from atomic to micron scale. *Nano Lett.* **16**, 5714–5718 (2016).
51. A. M. Hindeleh, R. Hosemann, Microparacrystals: The intermediate stage between crystalline and amorphous. *J. Mater. Sci.* **26**, 5127–5133 (1991).
52. J. H. E. Cartwright, A. G. Checa, B. Escribano, C. I. Sainz-Díaz, Spiral and target patterns in bivalve nacre manifest a natural excitable medium from layer growth of a biological liquid crystal. *Proc. Natl. Acad. Sci. U.S.A.* **106**, 10499–10504 (2009).
53. K. Wada, Spiral growth of nacre. *Nature* **211**, 1427 (1966).
54. N. Yao, A. Epstein, A. Akey, Crystal growth via spiral motion in abalone shell nacre. *J. Mater. Res.* **21**, 1939–1946 (2006).
55. S. Erland, P. E. Greenwood, Constructing 1/$\omega^\alpha$ noise from reversible Markov chains. *Phys. Rev. E Stat. Nonlin. Soft Matter Phys.* **76**, 031114 (2007).
56. S. Erland, P. E. Greenwood, L. M. Ward, "1/$f^\alpha$ noise" is equivalent to an eigenstructure power relation. *EPL* **95**, 60006 (2011).
57. P. Bak, C. Tang, K. Wiesenfeld, Self-organized criticality: An explanation of the 1/f noise. *Phys. Rev. Lett.* **59**, 381–384 (1987).
58. P. Bak, C. Tang, K. Wiesenfeld, Self-organized criticality. *Phys. Rev. A Gen. Phys.* **38**, 364–374 (1988).
59. J. Salman et al., Hyperspectral interference tomography of nacre. *Proc. Natl. Acad. Sci. U.S.A.* **118**, e2023623118 (2021).
60. C. Cajochen et al., Evidence that the lunar cycle influences human sleep. *Curr. Biol.* **23**, 1485–1488 (2013).
61. D. Tran et al., Field chronobiology of a molluscan bivalve: How the moon and sun cycles interact to drive oyster activity rhythms. *Chronobiol. Int.* **28**, 307–317 (2011).
62. M. Nakajima et al., Reconstitution of circadian oscillation of cyanobacterial KaiC phosphorylation in vitro. *Science* **308**, 414–415 (2005).

63. M. Scott, C. W. Gunderson, E. M. Mateescu, Z. Zhang, T. Hwa, Interdependence of cell growth and gene expression: Origins and consequences. *Science* **330**, 1099–1102 (2010).
64. L. Zhang *et al.*, Dissociation of circadian and circatidal timekeeping in the marine crustacean *Eurydice pulchra*. *Curr. Biol.* **23**, 1863–1873 (2013).
65. B. P. Tu, S. L. McKnight, Metabolic cycles as an underlying basis of biological oscillations. *Nat. Rev. Mol. Cell Biol.* **7**, 696–701 (2006).
66. H. V. Carey, M. T. Andrews, S. L. Martin, Mammalian hibernation: Cellular and molecular responses to depressed metabolism and low temperature. *Physiol. Rev.* **83**, 1153–1181 (2003).
67. K. B. Storey, J. M. Storey, Aestivation: Signaling and hypometabolism. *J. Exp. Biol.* **215**, 1425–1433 (2012).
68. P. Withers, S. Pedler, M. Guppy, Physiological adjustments during aestivation by the Australian land snail *Rhagada tescorum* (Mollusca: Pulmonata: Camaenidae). *Aust. J. Zool.* **45**, 599–611 (1997).
69. A. van der Weijden, M. Winkens, S. M. C. Schoenmakers, W. T. S. Huck, P. A. Korevaar, Autonomous mesoscale positioning emerging from myelin filament self-organization and Marangoni flows. *Nat. Commun.* **11**, 4800 (2020).
70. E. Karsenti, Self-organization in cell biology: A brief history. *Nat. Rev. Mol. Cell Biol.* **9**, 255–262 (2008).
71. E. Te Brinke *et al.*, Dissipative adaptation in driven self-assembly leading to self-dividing fibrils. *Nat. Nanotechnol.* **13**, 849–855 (2018).
72. M. Liu, Q. Li, F. Zhu, Threshold negative binomial autoregressive model. *Statistics* **53**, 1–25 (2019).
73. F. Qiu, H. Sridharan, Y. Chun, Spatial autoregressive model for population estimation at the census block level using LIDAR-derived building volume information. *Cartogr. Geogr. Inf. Sci.* **37**, 239–257 (2010).
74. R. J. Hayes, Methods for assessing whether change depends on initial value. *Stat. Med.* **7**, 915–927 (1988).
75. H. A. Hänni, A short review of the use of '*keshi*' as a term to describe pearls. *J. Gemmol.* **30**, 51–58 (2006).
76. A. Lin, M. A. Meyers, Growth and structure in abalone shell. *Mater. Sci. Eng. A* **390**, 27–41 (2005).
77. B. Runnegar, "The evolution of mineral skeletons" in *Origin, Evolution, and Modern Aspects of Biomineralization in Plants and Animals*, R. E. Crick, Ed. (Springer, Boston, MA, 1989), pp. 75–94.
78. J. Schwartz *et al.*, Removing stripes, scratches, and curtaining with nonrecoverable compressed sensing. *Microsc. Microanal.* **25**, 705–710 (2019).